\documentclass[twocolumn,twoside,slac_two]{revtex4}
\usepackage{graphicx}
\usepackage{fancyhdr}
\usepackage{graphics}
\usepackage{epstopdf}
\usepackage{textpos}
\begin{document}

%%%%%%%%%%%%%%%%%%%%%% WRITE THE TITLE HERE %%%%%%%%%%%%%%%%%%%
\title{\centering Parton Distribution Functions}
%%%%%%%%%%%%%%%%%%%%%% WRITE THE AUTHOR HERE %%%%%%%%%%%%%%%%%
\author{
\centering
\begin{center}
R. Pla\v cakyt\. e (on behalf of the H1 and ZEUS  Collaborations)    
\end{center}}
\affiliation{\centering Institut f\" ur Exp. Physik, Universit\" at Hamburg, Germany}
%%%%%%%%%%%%%%%%%%%%%% WRITE THE ABSTRACT HERE %%%%%%%%%%%%%%%%
\begin{abstract}
The parton distribution functions (PDFs) of the proton, a necessary input to almost all theory predictions
for hadron colliders, are reviewed in this document.
%An overview of the parton distribution functions (PDFs) is presented in this document. 
%he determination of PDFs and the main PDF fitting groups are introduced with the more detailed 
An introduction to the PDF determination by global analyses of the main PDF fitting groups
with an emphasis on HERA PDFs is presented.
%and the detailed description of the HERA PDFs is given.
Finally, theory predictions based on different PDFs are compared to some 
recent relevant LHC and TEVATRON measurements.
\end{abstract}

\maketitle
\thispagestyle{fancy}


\section{INTRODUCTION}
%Understanding of proton structure is esenitial for understanding of any physics
%process in hadron colliders.
A precise knowledge of the Parton Distribution Functions (PDFs) of the proton is essential in order to make predictions
for the Standard Model and beyond the Standard Model processes at hadron colliders.
%\\
%PDFs, $f_i(x,Q^2)$, represent the density of partons of flavour $i$ (quarks and gluons)
%in the proton with momentum fraction $x$ of the parton and $Q$ is the energy scale of the hard interaction.    
%Every cross section calculation is the convolution of the cross section at parton level and PDF’s.
%\\
The parton density function $f_i(x,Q^2)$ gives the probability of finding in the proton a parton 
of flavour $i$ (quarks or gluon) carrying a fraction $x$ of the proton momentum  with $Q$ being 
the energy scale of the hard interaction.
Cross sections are calculated by convoluting the parton level cross section with the PDFs.
%\\
%$f_i(x,Q^2)$, where $x$ is the momentum fraction of the parton involved in the hard process, 
%$Q$ is the energy scale of the hard interaction, and $i$ represents the parton flavour. 
Since QCD does not predict the parton content of the proton, the shapes of the PDFs are 
determined by a fit to data from experimental observables in various
processes, using the DGLAP evolution equation~\cite{DGLAPequations}.
%Thus, PDFs are essential for reliable predictions for new physics signals and their background
%cross sections at the LHC. 
\\
%Precise PDFs are also needed in order to maximise the discovery potential for new physics at the LHC. 
%The knowledge of proton PDFs primarily comes from the Deep Inelastic Scattering (DIS) data, e.g. HERA.
The knowledge of proton PDFs mainly comes from the Deep Inelastic Scattering (DIS) HERA, fixed target and 
TEVATRON data.
The recent precise TEVATRON and LHC data has a potential to improve constraints of the PDFs further.
%Since the LHC kinematic region is broader than currently explored,
%LHC has the opportunity to test QCD at very high- and low-$x$.
For example, through the $W$ lepton asymmetry the different quark contributions can be accessed 
and with inclusive jet productions it is possible to constrain the gluon PDF.
%\\
%In the next section a short overview of the PDF determination, current modern PDF sets and relevant 
%LHC processes are presented.

\section{PARTON DISTRIBUTION FUNCTIONS}

\subsection{Proton Structure and DIS}
\begin{figure}[!ht]
% \begin{minipage}[b]{0.5\linewidth} % A minipage that covers half the page
   \centering
   \includegraphics[width=5.6cm]{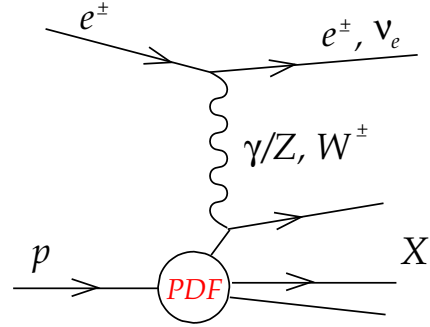}
% \end{minipage}
% \put (-7, 23) {$\Bigg X$} 
 \caption{\it Diagrams of neutral NC and charged CC current 
              deep inelastic scattering processes. The symbols denote the particles,
              the label "$X$" denotes the hadronic final state.} 
 \label{CC feynman diagrams}
\end{figure}

%Probability for a parton to carry the fraction x of proton momentum - Parton Distribution Function (PDF).
%Deep Inelastic Scattering (DIS) provides unique opportunity to study the structure of the proton.
%\\
Deep inelastic lepton nucleon scattering probes the structure of matter at
small distance scales.
At HERA (Hadron Elektron Ring Anlage), DESY, electrons (or positrons) were collided with protons
at centre-of-mass energies $\sqrt{s} = 225 - 318$~GeV.
%There are two deep inelastic $ep$ scattering processes measured at HERA: 
Two different deep inelastic $ep$ scattering processes are measured at HERA:
neutral current (NC), $ep~\to~eX$, and charged current (CC), $ep~\to~\nu X$.
%The exchanged particles between the electron and the quark
%in neutral current reactions are the photon ($\gamma$) and the boson $Z$ 
In neutral current reactions the interaction proceeds via the exchange of a photon 
or a Z boson while in charged current scattering a $W^\pm$ boson is exchanged 
(figure~\ref{CC feynman diagrams}).
%\\
%In the charged current scattering process a 
%charged boson $W^{\pm}$ is exchanged (figure~\ref{CC feynman diagrams}).
\\
The NC (and similarly CC) cross section can be expressed in terms of structure functions:
\begin{eqnarray}
  \nonumber
   \frac{d^2\sigma_{NC}^{e^{\pm} p}}{dxdQ^2}=\frac{2\pi\alpha^2}{xQ^4} 
     \big [ Y_{+} \tilde F_2^{\pm} \mp Y_{-}x \tilde F_3^{\pm} - y^2 \tilde F_L^{\pm} \big ],
 %\label{eq:NC not reduced}
\end{eqnarray}
where $Y_{\pm} = 1 \pm (1-y)^2$ with $y$ being the inelasticity. The structure function $\tilde F_2$
is the dominant contribution to the cross section, $x \tilde F_3$ is important at high $Q^2$ and $\tilde F_L$ is sizable 
only at high $y$. 
In the framework of perturbative QCD the structure functions are directly related to the 
parton distribution functions, i.e. in leading order (LO)  $F_2$ is the momentum sum of quark and anti-quark distributions, 
$F_2 \approx x \sum e^2_q (q+ \overline q)$, and $xF_3$ is related to their difference, 
    $xF_3 \approx x \sum 2e_q a_q (q- \overline q)$. At higher orders, terms related to the gluon density distribution
($\alpha_s g$) appear.
\\
In analogy to neutral currents, the inclusive CC $ep$ cross section can be expressed 
in terms of structure functions and in LO the $e^+p$ and $e^-p$ cross sections are sensitive to different quark 
densities:
\begin{eqnarray}
  \nonumber
    \begin{array}{rll}
   e^{+}:  & & \tilde \sigma_{CC}^{e^{+} p} = 
                x[\overline u +\overline c] + (1-y)^2 x[ d+s ]  \\
   e^{-}:  & & \tilde \sigma_{CC}^{e^{-} p} = 
                x[ u +c] + (1-y)^2 x[\overline d +\overline s ].
    \end{array}
\end{eqnarray}
 \vspace{0.07cm}
%\\       
%Measurements of NC and CC at HERA extended the kinematic regime in $Q^2$ by 
%more then two orders of magnitude 
%of the one achievable by the experiments which use stationary targets (e.g. SLAC).
%
The measurements of the NC and CC cross sections from HERA extend the kinematic regime 
in $Q^2$ by more than two orders of magnitude with respect to the fixed target experiments.

\subsection{Determination of PDFs}
%Although the parton distribution functions cannot be calculated, their $Q^2$ dependence
%is calculable in pQCD as described by the DGLAP
%(Dokshitzer-Gribov-Lipatov-Alterelli-Parisi) evolution equations~\cite{DGLAPequations}. 
%\\
%
%Generally, PDFs are obtained from fits of the DGLAP theory to 
%various sets of experimental measurements, many of which come from DIS experiments.
%The main parameters defining the PDF fit are: Order of the fit, input data, heavy flavour scheme, 
%assumptions in parametrisation, $\alpha_s$ and PDF correlation, treatment of uncertainties.
%
The general procedure to determine PDFs is as follows. Starting from a parameterisation of 
the non-perturbative PDFs at a low scale, either by making ad-hoc assumptions on their analytical 
form or by using the neural-net technology, fits to various sets of experimental data (mainly to DIS data)  
are performed within the DGLAP evolution scheme. The resulting PDFs depend on the choice of the input data, 
the order in which the perturbative QCD calculation is performed,
the assumptions about the PDFs, the treatment of heavy quarks,
the correlation between $\alpha_s$ and the PDFs and the treatment of the uncertainties.
Presently, the determination of PDFs 
is carried out by several groups, namely MSTW~\cite{MSTWpub}, CTEQ~\cite{CTEQpub}, 
NNPDF~\cite{NNPDFpub}, HERA\-PDF~\cite{herapdf10}, AB(K)M~\cite{ABKMpub} and GJR~\cite{GJRpub}.
In the following, a short description of each group with the relevant data sets used in PDF
fits is given.

The MSTW (Martin-Stirling-Thorne-Watt) PDFs are determined from a global analysis of hard-scattering 
data within the framework of leading-twist fixed-order collinear factorisation in 
the $\overline{\textrm{MS}}$
scheme~\cite{MSTWpub}. 
PDF sets in LO, NLO and NNLO are available. 
This determination is based on the HERA DIS data 
(with the exception of the latest combined HERA I data) as the main input for 
constraining quarks and gluon PDFs at low $x$, fixed target DIS data constraining quarks and gluons 
in the high-$x$ region, fixed target Drell-Yan (DY) data which help to constrain high-$x$ sea quarks, 
TEVATRON jet data contributing to high-$x$ gluon PDF 
and $W$, $Z$ data which provide an access to the different quark contributions.

The CTEQ PDFs~\cite{CTEQpub} are obtained by a global analysis of hard scattering data in the framework 
of general-mass perturbative QCD. This analysis is based on the same data mentioned above (CT10 also 
includes the recent combined HERA I data and more TEVATRON data). 
PDF sets in NLO and NNLO QCD are available.
%
%The Neural Net PDF (NNPDF) group~\cite{NNPDFpub} has a different approach to determine PDFs which
%includes the training a set of PDFs (i.e. minimisation of the $\chi^2$) parametrised by neural networks 
%on MC replicas given by the experimental data. It also includes all kind of experimental data
%in the fit and provides LO, NLO and NNLO PDFs.
%

The Neural Net PDF (NNPDF) group~\cite{NNPDFpub} parametrise non-perturbative PDFs
by training a neural network (i.e. minimising the $\chi^2$) on MC replicas of the experimental data.
With such a method NNPDFs are free of assumptions made by other groups (however there is
some ambiguity from the procedure itself).
It also includes almost all above mentioned experimental data in the fit and provides LO, NLO and NNLO PDFs.

%
%allows to use conventional $\chi^2$ tolerance $\Delta \chi^2 = 1$ and thus does not require nuclear correction
%factors which are necessary for fixed target data. 
The HERAPDFs~\cite{herapdf10} are based only on HERA DIS data which 
allows the usage of the conventional $\chi^2$ tolerance of $\Delta \chi^2=1$. 
Since this analysis is solely based on $ep$ data, the PDFs do not depend on the approach for nuclear 
corrections needed for fixed target data.
The HERAPDF has been extracted in NLO and NNLO. 
More details about HERA PDFs are given in the next section.       
        
The ABKM~\cite{ABKMpub} group provides NLO and NNLO PDFs
%using a consistent scheme for matching the heavy quark structure functions. 
based on the inclusive NC DIS world data from HERA and from fixed-target experiments.
They also provide the strong coupling constant $\alpha_s$ in NNLO. 

\begin{figure}[!ht]
   \centering
  % \includegraphics[width=8cm]{w+w-lhc7nlo68err.eps}
  % \caption{\it The NLO $W^+$ and $W^-$ cross section predictions using different PDFs at the LHC
   \includegraphics[width=8cm]{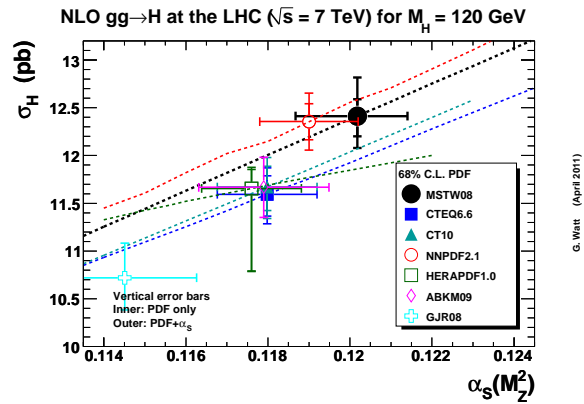}
 \caption{\it NLO Higgs cross section predictions ($M_H = 120$ GeV) using different PDFs at the LHC
              with $\sqrt s = 7$ TeV.}
 %\label{wpwmfig}
 \label{higgsfig}
\end{figure}

The GJR~\cite{GJRpub} group uses a more restrictive “dynamical” approach in the PDFs determination, 
i.e. a dynamical model is used in the evolution of the input distributions starting at very low $Q^2$.
%starting evolution at very low $Q^2$ where the nucleon only consists of valence quarks (provides small uncertainty at low $x$).
Fits to DIS, DY and TEVATRON jet data provide LO, NLO and NNLO PDF sets.
%both in the FFNS and in the VFNS. 
\\
\\        
%The main parameters defining the PDF fit are: Order of the fit, input data, heavy flavour scheme, 
%assumptions in parametrisation, $\alpha_s$ and PDF correlation, treatment of uncertainties.
The large number of PDF parameters and 
their treatment in the fitting procedure within the different groups results in differences of the PDFs provided.
%their different treatment within the group
%in the PDF fitting procedure cause also differences in the provided PDFs sets.
%There are differences in how these parameters are treated in the fitting
%procedure between the groups thus causing differences in PDFs sets.
In order to study these differences, 
%(and give a respons to the LHC Higgs group ?)        
a benchmarking exercise is being carried out by the PDF4LHC working group formed by
the members of the PDF fitting groups mentioned above. As an example of this exercise,       
the NLO prediction of the Higgs cross section ($M_H = 120$~GeV) for the LHC is shown in 
figure~\ref{higgsfig} for different PDF sets as a function of $\alpha_s$. 
It can be seen that the different PDF groups are using different default values of $\alpha_s$ 
and the cross section predictions differ by a few percent when using the default values of $\alpha_s$ 
as well as when using the same value for $\alpha_s$. 
Furtehermore, the different predictions have different uncertainties  
%
%the NLO Higgs ($M_H = 120$~GeV) cross section predictions at the LHC as a function of $\alpha_s$
%using different PDFs are illustrated in figure~\ref{higgsfig}. 
%As can be seen from this figure, first, the default $\alpha_s$ value is used in different PDF fitting groups, 
%second, the cross section predictions differ by few percent at the default as well as the same $\alpha_s$ value.
%Finally, it can be seen that different predictions yield different uncertainties 
(e.g. HERA PDF has asymmetric uncertainties due
to considered parametrisation variation as explained in the next section). \\
Further comparisons and an overview of the PDF4LHC working group activities can be found in~\cite{pdf4lhc}. 
%\\
%\vspace{-0.05cm}

% figures should be put into the text as floats.
% Use the graphics or graphicx packages (distributed with LaTeX2e)
% and the \includegraphics macro defined in those packages.
% See the LaTeX Graphics Companion by Michel Goosens, Sebastian Rahtz,
% entire page. There is no need to do explicit centering.

% \begin{figure}
% \end{figure}

% Surround figure environment with turnpage environment for landscape
% figure
% \begin{figure}
% \end{figure}

%\begin{figure*}[t]
%\centering
%%\includegraphics[width=135mm]{JACpic2.eps}
%\caption{Example of full width figure.} \label{JACpic2-f1}
%\end{figure*}

\subsection{HERA PDFs}
\begin{figure}[!ht]
   \centering
   \includegraphics[width=8cm]{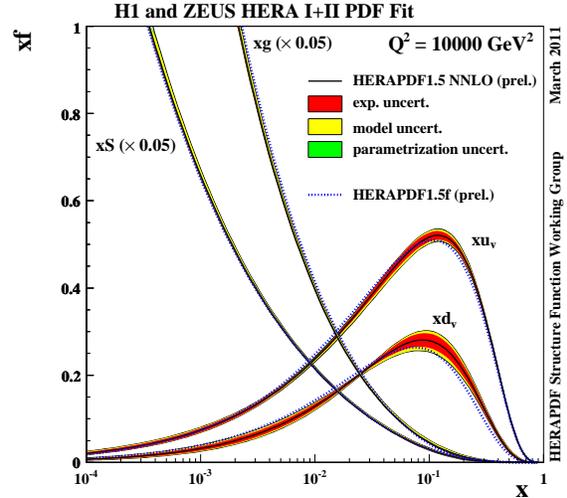}
 \caption{\it The parton distribution functions from HERAPDF1.5 NNLO
             at $Q^2$ = 10000 GeV$^2$, i.e. a region relevant for the hadron colliders TEVATRON and LHC. 
              The gluon and sea distributions are scaled
             down by a factor 20. The experimental, model and parametrisation uncertainties are shown
             separately. For comparison, the central values of HERAPDF1.0 NNLO are also shown.}
 \label{herapdf15nnlofig}
\end{figure}

In the HERAPDF1.0 fit~\cite{herapdf10}, the combined set on NC and CC $ep$ 
inclusive cross-sections for the first running period of HERA (HERA I) is used. 
%input for a next-to-leading order (NLO) QCD PDF fit, called HERAPDF1.0~\cite{herapdf10}. 
The full statistics HERA inclusive CC and NC data are used for NLO and NNLO QCD fits resulting
in HERAPDF1.5~\cite{herapdf15}.
The same formalism, model and paramatrisation assumptions as in the HERAPDF1.0 
are used in the HERA\-PDF1.5(NLO) fit.
%(in a later variants of HERAPDFs a more flexible parametrisation is assumed).
%and are shortly described below. 
\\
\\
The QCD predictions for the structure functions are obtained by solving the DGLAP evolution equations 
at NLO (or NNLO)
in the $\overline{MS}$ scheme with the renormalisation and factorisation scales chosen to be $Q^2$.
The DGLAP equations yield the PDFs at all values of $Q^2$ above the input scale $Q^2_0$ at which they are 
parametrised as a functions of $x$.
The starting scale $Q^2_0$
%, at which the PDFs are provided as functions of $x$ in the fit, 
is chosen to be $1.9$ GeV$^2$ such that the starting scale is below the charm mass threshold.
\\
The QCD predictions for the structure functions are obtained by convolution of the PDFs
with the NLO coefficient functions calculated using the general mass variable favour number 
RT scheme~\cite{RTref}.
\\
For the parametrisation of PDFs at the input scale the generic form $xf(x)=Ax^B(1-x)^C(1+Ex^2)$ is used.        
The parametrised PDFs are the gluon distribution $xg$, the valence quark distributions $xu_v$, $xd_v$,
and the u-type and d-type anti-quark distributions $x\bar U$, $x\bar D$.
At the starting scale $Q^2_0 = 1.9$ GeV$^2$ $ x\bar U~=~x\bar u$ 
and $x\bar D = x\bar d + x\bar s$.
The central fit parametrisation is:
% -------------------- param of HERAPDF
%\begin{flushleft}
  \begin{eqnarray}
     \nonumber
      xg(x)      & = & A_gx^{B_g}(1-x)^{C_g},  \\
     \nonumber
      xu_v(x)    & = & A_{u_v}x^{B_{u_v}}(1-x)^{C_{u_v}}(1+E_{u_v}x^2), \\
     \nonumber
      xd_v(x)    & = & A_{d_v}x^{B_{d_v}}(1-x)^{C_{d_v}}, \\
     \nonumber
      x\bar U(x) & = & A_{\bar U}x^{B_{\bar U}}(1-x^{C_{\bar U}}), \\
     \nonumber
      x\bar D(x) & = & A_{\bar D}x^{B_{\bar D}}(1-x^{C_{\bar D}}).
  \end{eqnarray}
%\end{flushleft}
The normalisation parameters $A$ are constrained by the quark number sum-rules and momentum sum-rule,
extra constrains for small-$x$ behaviour of $d-$ and $u-$type quarks $B_{u_v}~=~B_{d_v}$, $B_{\bar U}~=~B_{\bar D}$
and $A_{\bar U}~=~A_{\bar D}(1-f_s)$ ($f_s$ is the strange quark distribution) which ensures that 
$x\bar u \rightarrow x\bar d$  as $x \rightarrow 0$.
\\
\begin{figure}[!ht]
   \centering
   \includegraphics[width=8cm]{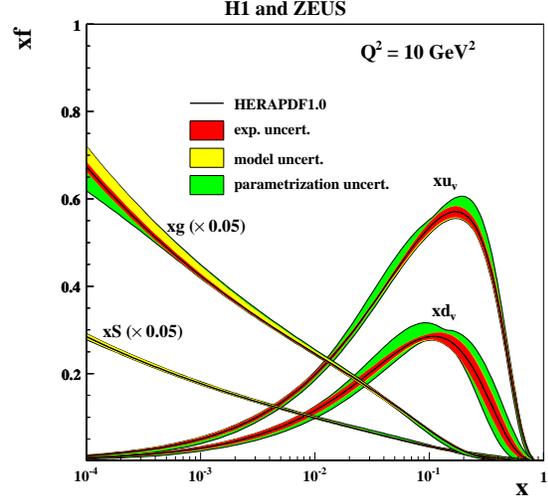}
 \caption{\it The parton distribution functions from HERAPDF1.0
             at $Q^2$ = 10 GeV$^2$. The gluon and sea distributions are scaled
             down by a factor of 20. The experimental, model and parametrisation uncertainties are shown
             separately.}
 \label{herapdf10fig}
\end{figure}
\\
The break-up of the HERA PDFs into different flavours is illustrated in figure~\ref{herapdf10fig}.       
Model uncertainties (shown as yellow bands in the figure) of the central fit solution is evaluated
by varying the input assumptions: $Q^2_{min}$, $f_s$, mass of heavy quarks $m_C$ and $m_B$.
Parametrisation uncertainties (green band) is formed by an envelope of the maximal deviation
from the central fit varying parametrisation assumptions in the fit and therefore has an
asymmetric shape. The determination of parameterisation uncertainties are unique to HERAPDFs.
\\
%Recently a new set of HERAPDFs were released which are based on full HERA statistics inclusive sample 
%at NLO as well as NNLO order (HERAPDF1.5NLO and HERAPDF1.5NNLO). 
%\\        
An example of the parton distribution functions from HERAPDF1.5 at NNLO is shown in figure~\ref{herapdf15nnlofig}.       
HERAPDF1.5NLO and NNLO sets are the recommended HERA PDFs to be used for the predictions of processes at LHC. 

\subsection{Comparisons to recent LHC and TEVATRON results}

\begin{figure}[!ht]
   \centering
   \includegraphics[width=8.0cm]{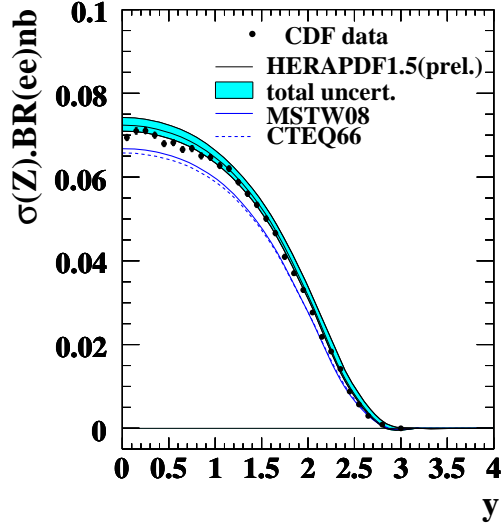}
 \caption{\it $Z$ boson cross section times the branching ratio to electrons as a function of the rapidity $y$, 
     as measured by the CDF collaboration. The cross 
                  sections are compared to the theory predictions based on three different PDFs. The total PDF uncertainties
                  on the prediction on HERA\-PDF1.5 is shown as a blue band.  }
 \label{cdfz0fig}
\end{figure}

\begin{figure}[!ht]
   \centering
   \includegraphics[width=8.1cm]{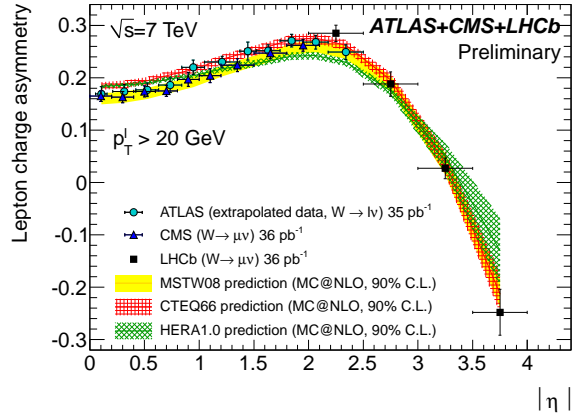}
 \caption{\it The W lepton charge asymmetry measured by ATLAS, CMS and LHCb (symbols) compared to the 
              predictions based on different PDFs shown as lines with uncertainty bands. }
 \label{atlascmslhcbfig}
\end{figure}

\begin{figure}[!ht]
   \centering
   \includegraphics[width=8.1cm]{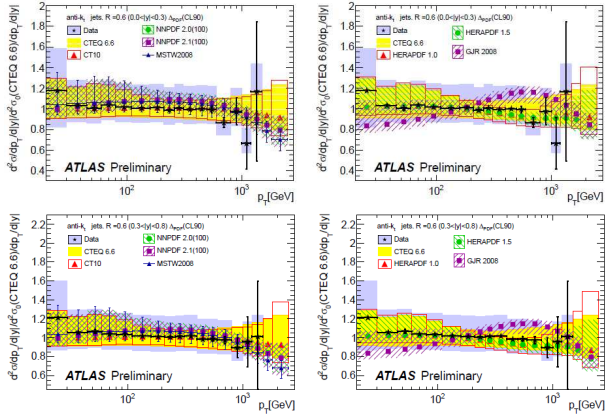}
 \caption{\it Inclusive double-differential jet cross sections as a function of jet $P_T$ in different rapidity regions
              for jets identified using the anti-k$_t$ algorithm with $R = 0.6$ (ATLAS). 
              The plot shows the ratio of the data to the theoretical predictions calculated with different PDFs (CTEQ 6.6 
              is used as a reference).}
 \label{atlasjetfig}
\end{figure}

\begin{figure}[!ht]
   \centering
   \includegraphics[width=7.9cm]{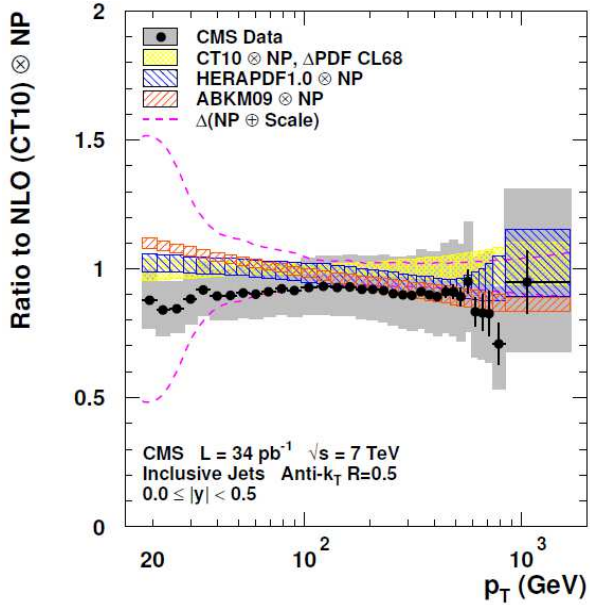}
 \caption{\it The ratio of inclusive jet cross section (CMS) for $|y| < 0.5$ as a function of jet $P_T$ 
              to NLO predictions based on the CT10 PDF set.
              Additional predictions are shown using HERAPDF1.0 and ABKM09 PDFs. PDF uncertainties are
              shown as coloured bands. }
 \label{cmsjetfig}
\end{figure}

%An overview of the most recent comparisons of theory predictions obtained with different PDFs
%to the latest TEVATRON and LHC data is given in this section.
%\\
The prediction of the $Z$ boson rapidity distribution, based on three different PDFs, are compared to the CDF 
measurement in figure~\ref{cdfz0fig}.
%Good agreement with predictions obtained using HERAPDF1.5 
%as well as other  PDFs is observed.
%\\
%\\
\begin{figure}[!ht]
   \centering
   \includegraphics[width=8.0cm]{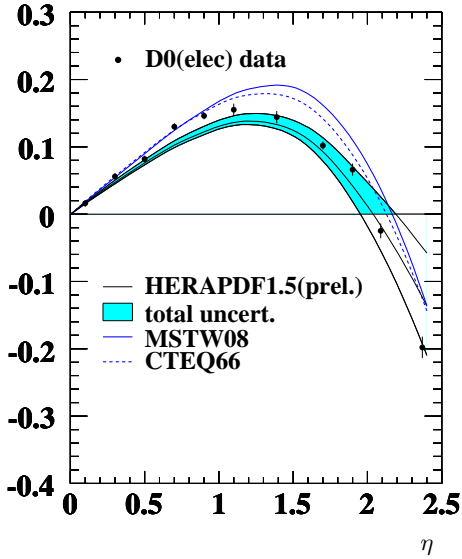}
%   \put (-35, 23)  { \textbf $ \eta $ } 
   \put (-35, 23)  {\textbf{ $\eta$} } 
 \caption{\it The electron charge asymmetry as a function of the lepon pseudorapidity $\eta$, as measured by the D0 collaboration. 
                  The measurement is  compared to the theory predictions based on three different PDFs. The total PDF uncertainties
                  on the prediction on HERA\-PDF1.5 is shown as a blue band.  }
 \label{d0elecasymfig}
\end{figure}
%\\
The predictions of the $W$ boson lepton asymmetry, based on the same PDFs
%Similarly, the prediction of $W$ boson electron asymmetry based same PDFs 
are compared to the 
D0 measurement in figure~\ref{d0elecasymfig}. As can be seen from these figures,
the predictions with the HERAPDF sets are in good agreement with data 
and the agreement is comparable to that obtained for the CTEQ66 and MRSTW08 PDF sets.
\\
\\
%After about two years of successful LHC running there are many new physics results being released by LHC experiments.
Recently many new physics results are released by the LHC experiments.
The high precision LHC data can be used not only to compare with theory predictions 
to differentiate between different PDF sets but can also be used to put constraints on them instead.
For example, the $W$ lepton charge asymmetry $A$ can help to constrain the $u$ to $d$ quark ratio 
($A_w = (W^+ - W^-) /  (W^+ + W^-)$ can be approximated to the ratio 
$A_W \approx (u_v - d_v) / (u_v + d_v + 2u_{sea})$ and thus is sensitive to the valence $u$ and $d$ quark ratio), 
the top quark
measurement would have impact on gluons in the high-$x$ region, inclusive jet measurements can be used to 
improve constrains on the gluon
(the~$gg$ initial state has dominant contribution to the cross section at the low jet transverse momenta,~$P_T$).
%\\
\\
%but data becomes precise enough to start to discriminate between different predictions.
The $W$ boson lepton asymmetry as measured by ATLAS, CMS and LHCb is shown in figure~\ref{atlascmslhcbfig}.       
%As mentioned before, the asymmetry $A_w = (W^+ - W^-) /  (W^+ + W^-)$ can be approximated to the ratio
%$A_W \approx (u_v - d_v) / (u_v + d_v + 2u_{sea})$
%and thus is sensitive to the valence $u$ and $d$ quark ratio. 
The measured asymmetry is compared to predictions obtained with MSTW08, CTEQ66 and HERAPDF1.0 PDFs. 
The PDF uncertainties are shown as bands for each prediction.
There is already a more precise $W$ asymmetry preliminary measurement available from the CMS 
collaboration~\cite{cmsprelimw}.
\\
\\
The inclusive jet cross section is sensitive to the gluon density and can help to constrain it.
In figure~\ref{atlasjetfig} the inclusive jet cross section measured by the ATLAS collaboration is shown 
as a function of the jet transverse momentum in different rapidity regions. 
The theoretical predictions based on different PDF sets are describing the data equally well.
\\
A similar measurement was performed by the CMS collaboration~\cite{cmsjets}.
Figure~\ref{cmsjetfig} shows   
the inclusive jet cross section measured by CMS as a function of jet transverse
momentum in the rapidity bin $0.0~\leq~y~\leq~0.5$. 
The measurement is compared to predictions calculated using different PDFs
which are agreeing similarly well with the data as in the case of ATLAS measurement. 
\vspace{0.1cm}
%From the figure it can be seen that this measurement has similar level of agreement with theory predictions 
%obtained with different PDFs.
%\\

\subsection{Recent Updates of HERAPDFs}
\begin{figure}[!ht]
   \centering
   \includegraphics[width=7.8cm]{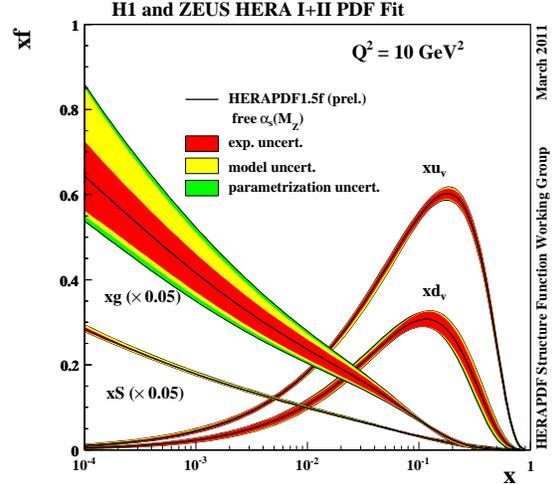}
 \caption{\it The parton distribution functions from HERAPDF1.5
             at $Q^2$~=~10~GeV$^2$ with $\alpha_s (M_Z)$ treated as a free parameter in the fit.}
 \label{herapdfjet01fig}
\end{figure}

\begin{figure}[!ht]
   \centering
   \includegraphics[width=7.8cm]{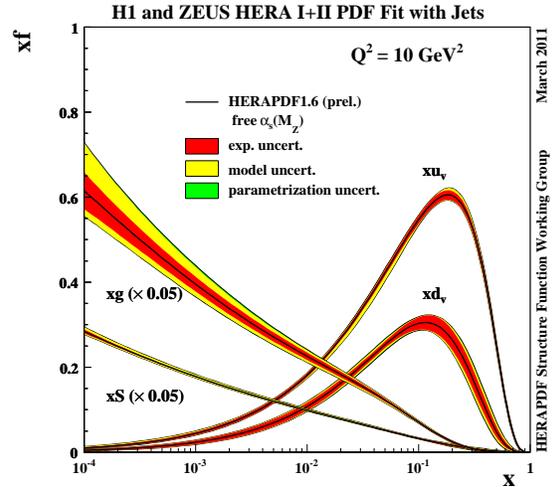}
 \caption{\it The parton distribution functions from HERAPDF1.6 (with HERA jet data included in the fit)
             at $Q^2$ = 10 GeV$^2$ with $\alpha_s (M_Z)$ treated as a free parameter in the fit.}
 \label{herapdfjet02fig}
\end{figure}

\begin{figure}[!ht]
   \centering
   \includegraphics[width=8cm]{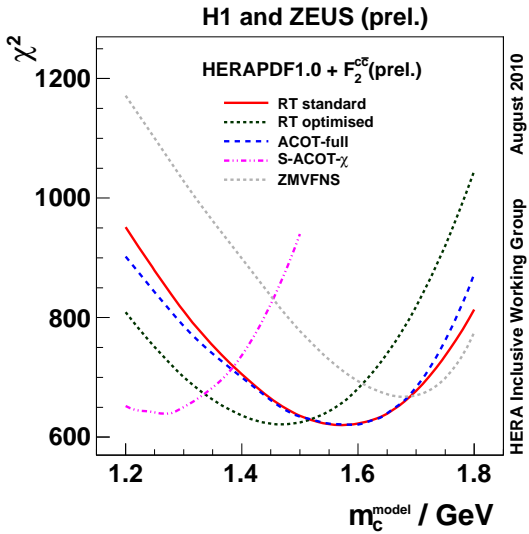}
 \caption{\it  Comparison of the $\chi^2$ of HERA I + F$_2^{\bar cc}$ fits
             using different heavy flavour schemes represented as lines of different styles.}
 \label{herapdfcfig}
\end{figure}

\begin{figure}[!ht]
   \centering
   \includegraphics[width=8cm]{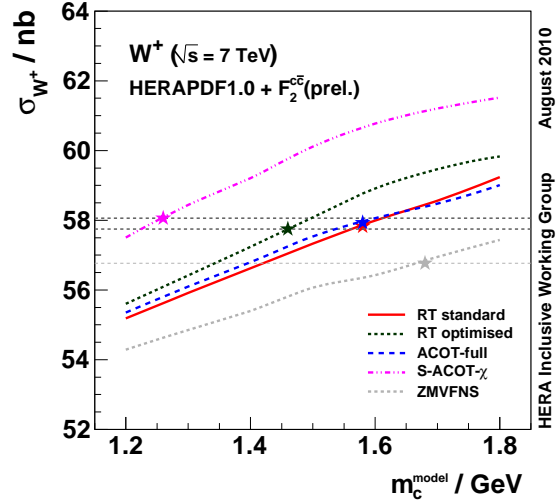}
 \caption{\it $W^{+}$ production cross section $\sigma_{W^{+}}$ at the LHC for $\sqrt s = $ 7 TeV
     as a function of $m_c^{mod}$. The lines show predictions for different VFN schemes as indicated by the legend.
     The stars show the predictions obtained with the optimal value of $m_c^{mod}$ used in a given scheme.
     The thick dashed horizontal lines indicate the range of $\sigma_{W^{+}}$, determined for 
     $m_c^{mod}$ = $m_c^{mod}$ (opt).} 
 \label{herapdfc_lhcfig}
\end{figure}

The HERAPDF1.5 analysis has been extended to include HERA inclusive jet data. The new
PDF set with jet data is called HERAPDF1.6~\cite{herapdf16}.
%
%%As mentioned before, the jet cross sections constrain gluon density and thus can improve 
%%knowledge of gluon PDF in large-$x$ region. 
The jet data are sensitive to gluons and $\alpha_s$ and therefore  
reduce the correlation between the gluon density and $\alpha_s$ in the PDF fits.
% Karin: the jet data mainly constrain the gluon in the region $0.01<x<0.1$. 
%When fixing alphas the pdfs are not improved, but if you look carefully you see a 
%slight change in the gluon pdf for 0.01<x<0.1. The main impact the jet data are 
%making is the possibility to disentangle the effect from the gluon and alphas.
%
The impact of the jet data can be seen if $\alpha_s (M_Z)$ 
is treated as a free parameter in the fit
(see figure~\ref{herapdfjet01fig} and~\ref{herapdfjet02fig}).
In case of fitting simultaneously PDFs and $\alpha_s$ to the inclusive DIS
%In addition, there is a strong correlation between the low-$x$ shape of the gluon PDF 
%and $\alpha_s (MZ)$. 
%In this case the uncertainty on the gluon PDF at low-$x$ becomes very large but, 
data the uncertainties on the gluon PDF becomes large at low $x$ but
as soon as the jet data
are included (figure~\ref{herapdfjet02fig}), the correlation between the gluon PDF and $\alpha_s (M_Z)$
is reduced, resulting 
in significantly reduced uncertainties on the gluon PDF 
similar to the case of using inclusive DIS data only and fixing $\alpha_s (M_Z)$.
In the fit with jet data the value of $\alpha_s$ was extracted:
\\
$\alpha_s (M_Z)~=~0.1202~\pm~0.0013$(exp)~$\pm~0.0007$(mod/param) $\pm 0.0012$(hadronisation) $^{+0.0045}_{-0.0036}$(scale) ~\cite{herapdf16}.
\\
More details about QCD analysis with jets at
HERA were presented in the talk by T.~Schoerner-Sadenius~\cite{ThomasTalk}.
\\        
\\
The effect of including the HERA charm data in the QCD PDF fits was also studied at HERA.
In this study, the charm data are used together with different implementations of 
the variable flavour number schemes (VFNS), which have different approaches for the interpolation 
function and counting of orders in $\alpha_s$. Different VFN schemes can have different 
impacts on the charm contribution to the sea quark distribution
%impact to the charm in the sea distribution 
and thus affect the composition of $x \overline U(x)$ from the $x\overline u(x)$
and the $x\overline c(x)$ contributions.
Therefore the accuracy of the charm data influence the uncertainties on the $W^{\pm}$ and $Z$ 
cross section predictions at LHC.
%via the compensation of $u$ vs $c$ quarks which are related via structure function $F_2$.
Figure~\ref{herapdfcfig} summarises the study by showing the charm quark mass $m_c^{mod}$ (which is an 
effective parameter in the fit) scanning results for all schemes used in the fits. 
It is interesting to observe that first, 
different schemes have different optimal charm quark mass 
parameter $m_c^{mod}$, and second, the $\chi^2$ minimum values
are comparable for all schemes despite of different optimal values of $m_c^{mod}$.
%\\
\\
The PDFs with optimal $m_c^{mod}$  were then propagated to the calculation of $W^{\pm}$ and $Z$
boson cross section predictions for the LHC. As an example, 
the $W^{+}$ cross section as a function of $m_c^{mod}$ for the different schemes is shown in
figure~\ref{herapdfc_lhcfig}. 
Good agreement between these predictions is observed at optimal $m_c^{mod}$
which results in a reduction of the uncertainties due to the heavy flavour treatment to below~1.0$\%$. 
More details about
\begin{figure}[!ht]
   \centering
   \includegraphics[width=8cm]{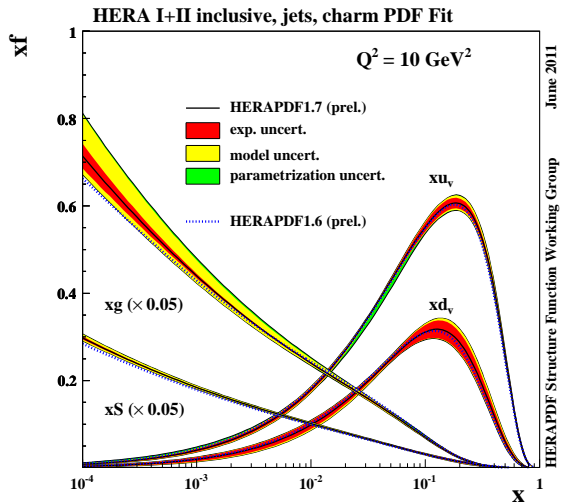}
 \caption{\it The parton distribution functions from HERAPDF1.7
             at $Q^2$ = 10 GeV$^2$. The gluon and sea distributions are scaled
             down by a factor of 20. The experimental, model and parametrisation uncertainties are shown
             separately. For comparison, the central values of HERAPDF1.6 are also shown.}
 \label{herapdf17fig}
\end{figure}
the study can be found in~\cite{charmnote}.
%
%Fitting different HERA data - HERAPDF1.7 - consitency check.
\\ \\
Finally, a QCD fit was performed which includes all currently available HERA NC and CC inclusive, charm,
low energy and jet data (HERAPDF1.7). 
%
%The jet data prefers the higher value of $\alpha_s$ than used in the default HERA PDFs (from 0.1176 to 0.119)
%as suggested previously by the HERAPDF1.6, and charm data prefers $m_C$ value 1.5 together with the RT-optimised heavy 
%flavour scheme following the preference of the previous fit with the charm data. 
%
This fit prefers the same $\alpha_s$ value as obtained with a fit using only inclusive and jet data.
Similarly, the same optimal $m_C$ value was found in the HERAPDF1.7 fit as that obtained 
in a fit when only inclusive and charm data are fitted.
The break-up of the HERAPDF1.7 PDFs into different flavours is illustrated in figure~\ref{herapdf17fig}.

\section{SUMMARY}

Parton distribution functions of the proton are essential for any cross section prediction at hadron colliders.    
PDFs are generally determined by global QCD analyses fitting various sets of experimental data   
with the main knowledge about the proton structure coming from DIS experiments.
Six modern PDF sets are currently available and shortly reviewed here: MSTW, CTEQ, NNPDF, HERAPDF, AB(K)M and GJR.
The detailed description of the HERA PDFs fit, parametrisation and model assumptions are presented. 
Recent developments in the HERAPDF fits include the QCD analysis of HERA inclusive and jet data, 
of HERA inclusive and charm data and an analysis in which the consistency
%Recent developments in HERA PDFs include QCD analysis with HERA jet data, charm measurements and 
%analysis where the consistency 
is checked fitting HERA inclusive data together with jet and charm measurements. \\
Finally, comparisons of theory predictions with various PDF sets to the latest LHC and TEVATRON data are shown.


%\begin{acknowledgments}
%The authors wish to thank JACoW for their guidance in preparing
%this template.
%
%Work supported by Department of Energy contract DE-AC03-76SF00515.
%\end{acknowledgments}

\bigskip % extra skip inserted
\bigskip % extra skip inserted
\bigskip % extra skip inserted
\bigskip % extra skip inserted
\bigskip % extra skip inserted
\bigskip % extra skip inserted
\bigskip % extra skip inserted
\bigskip % extra skip inserted
\bigskip % extra skip inserted
\bigskip % extra skip inserted
\bigskip % extra skip inserted
\bigskip % extra skip inserted
\bigskip % extra skip inserted
%% Create the reference section using BibTeX:
\bibliography{basename of .bib file}

\begin{thebibliography}{9}   % Use for  1-9  references
%%\begin{thebibliography}{99} % Use for 10-99 references
%
%%\bibitem{accelconf-ref}
%%http://www.cern.ch/accelconf
%
%\bibitem{exampl-ref}
%%%%%%%%%\bibitem{tom} T.~Junk, Nucl. Instrum. Meth. A {\bf 434}, 435 (1999)
%%\bibitem{templates-ref}
%%http://www.cern.ch/accelconf/templates.html

\bibitem{DGLAPequations}
V.N. Gribov, L.N.~Lipatov, Sov. J. Nucl. Phys.~{\bf 15}, 438, 675 (1972), \\
L.N. Lipatov, Sov. J. Nucl. Phys.~{\bf 20}, 94 (1975), \\
G. Altarelli, G. Parisi, Nucl. Phys. B{\bf 126}, 298 (1977), \\
Yu. L. Dokshitzer, Sov. Phys. JETP {\bf 46}, 641 (1977), \\
G. Curci, W. Furmanski, and R. Petronzio, Nucl.Phys. B{\bf 175}, 27 (1980), \\
S. Moch, J. Vermaseren, and A.~Vogt, Nucl.Phys. B{\bf 688}, 101 (2004) [arXiv:hep-ph/0403192], \\
A. Vogt, S. Moch, and J.~Vermaseren, Nucl.Phys. B{\bf 691}, 129 (2004) [arXiv:hep-ph/0404111].

\bibitem{MSTWpub}
A.D. Martin, W.J. Stirling, R.S. Thorne, G.~Watt, Eur. Phys. J. C{\bf 63}, 189 (2009) [arXiv:0901.0002].

\bibitem{CTEQpub}
P. M. Nadolsky et al, Phys. Rev. D{\bf 78}, 013004 (2008) [arXiv:0802.0007].

\bibitem{NNPDFpub}
Richard D. Ball et al., Nucl. Phys. B{\bf 809}, 1 (2009) [arXiv:0808.1231v4],  \\
Richard D. Ball et al., Nucl.Phys. B{\bf 838}, 136 (2010) [arXiv:1002.4407]. 

\bibitem{herapdf10}
F. Aaron at al. $[$H1 and ZEUS Collaborations$]$, JHEP B{\bf 1001}, 109 (2010) [0911.0884].

\bibitem{ABKMpub}
S. Alekhin, J. Bl\" umlein, S. Klein, S. Moch, Phys. Rev. D{\bf 81}, 014032 (2010) [arXiv:0908.2766].

\bibitem{GJRpub}
M. Gl\" uck, P. Jimenez-Delgado, E.~Reya, Eur.~Phys. J. C{\bf 53}, 355 (2008) [arXiv:0709.0614], \\
M. Gl\" uck, P. Jimenez-Delgado, E.~Reya, C.~Schuck, Phys. Lett. B{\bf 664}, 133 (2008) [arXiv:0801.3618].

\bibitem{pdf4lhc}
URL {\tiny https://wiki.terascale.de/index.php?title=PDF4LHC$\_$WIKI}.

\bibitem{herapdf15}
$[$H1 and ZEUS Collaborations$]$, H1prelim-10-142, ZEUS-prel-10-018, \\
$[$H1 and ZEUS Collaborations$]$, H1prelim-11-042, ZEUS-prel-11-002.

\bibitem{RTref}
R. S. Thorne and R. G.~Roberts, Phys. Rev. D{\bf 57}, 6871 (1998) [hep-ph/9709442]. 

\bibitem{cmsprelimw}
$[$CMS Collaboration$]$, CMS-PAS-EWK-11-005.

\bibitem{cmsjets}
$[$CMS Collaboration$]$, CMS-NOTE-2011-004, CERN-CMS-NOTE-2011-004.

\bibitem{herapdf16}
[H1 and ZEUS Collaborations], H1prelim-11-034, ZEUS-prel-11-001.

\bibitem{ThomasTalk}
Thomas Sch\" orner-Sadenius, "Measurement of $\alpha_s$ in DIS", these proceedings.

\bibitem{charmnote}
[H1 and ZEUS Collaborations], H1prelim-10-143, ZEUS-prel-10-019.

%
\end{thebibliography}

\end{document}